# Agentic AI, Medical Morality, and the Transformation of the Patient-Physician Relationship

Robert Ranisch[a]*     Sabine Salloch[b]

[a] Faculty of Health Sciences, University of Potsdam, Potsdam, Germany

[b] Institute for Ethics, History and Philosophy of Medicine, Hannover Medical School, Hanover, Germany

* Corresponding author: Robert Ranisch, Junior Professorship for Medical Ethics with a Focus on Digitization, Faculty of Health Sciences Brandenburg, University of Potsdam, Am Mühlenberg 9, 14476, Potsdam, Golm, Germany, ranisch@uni-potsdam.de

February 18, 2026

## Abstract

The emergence of agentic AI marks a new phase in the digital transformation of healthcare. Distinct from conventional generative AI, agentic AI systems are capable of autonomous, goal-directed actions and complex task coordination. They promise to support or even collaborate with clinicians and patients in increasingly independent ways. While agentic AI raises familiar moral concerns regarding safety, accountability, and bias, this article focuses on a less explored dimension: its capacity to transform the moral fabric of healthcare itself. Drawing on the framework of techno-moral change and the three domains of decision, relation and perception, we investigate how agentic AI might reshape the patient-physician relationship and reconfigure core concepts of medical morality. We argue that these shifts, while not fully predictable, demand ethical attention before widespread deployment. Ultimately, the paper calls for integrating ethical foresight into the design and use of agentic AI.

## 1. The rise of agentic AI in healthcare

The recent introduction of generative AI, particularly large language models (LLMs), in clinical settings has been heralded as a major advancement in making effective AI tools for healthcare provision feasible (Lee et al. 2023). Many promising applications have been proposed, ranging from diagnosis and treatment planning to administrative tasks and medical education (Haltaufderheide and Ranisch 2024; Busch et al. 2025; Vrdoljak et al. 2025).

Alongside the rise of generative AI, some have begun to anticipate a "new era of agentic medicine" (Zou and Topol 2025). Agentic Artificial Intelligence (AI) is increasingly considered the next major paradigm in an evolving field of medical AI (Moritz et al. 2025; Taylor 2025; Xu et al. 2025). While the term AI agent is sometimes used loosely to describe these systems, there is currently no uniform terminology in the literature (Gorenshtein et al. 2025). Some authors use "AI agent" and "agentic AI" interchangeably, whereas others distinguish them more clearly. Sapkota et al. (2026), for example, define "AI agents" as modular systems that are driven and enabled by LLMs (and potentially Large Image Models, LIMs) to perform task-specific automation. "Agentic AI", by contrast, they see as characterized by "multi-agent collaboration, dynamic task decomposition, persistent memory, and coordinated autonomy." (Sapkota et al. 2026, p. 1). In this article, we distinguish between single AI agents – systems that perform specific tasks – and agentic AI, which refers to orchestrated networks of such agents capable of pursuing complex, multi-step goals. Our main focus will be on the introduction of agentic AI networks in healthcare.

The functionality of agentic AI in many respects exceeds that of single LLMs. LLMs are designed to process and produce human-like text by understanding and generating written outputs efficiently across various contexts. In healthcare, LLMs are primarily used for tasks involving natural language processing, such as automating documentation, translating medical jargon for better patient understanding or assisting with the synthesis of medical literature (Haltaufderheide and Ranisch 2024; Leiser et al. 2025; Li et al. 2025). Such scenarios require detailed communication and information retrieval, yet they lack autonomous decision-making capabilities.

Agentic AI, by contrast, embodies a more advanced paradigm with capabilities that extend beyond reactive language processing. Unlike LLMs, which serve primarily assistive and communicative functions, agentic AI is able to execute goal-directed tasks in the digital or physical world, interact with external tools, and coordinate multi-step processes with varying degrees of autonomy. This potentially allows such AI systems to determine actions related to diagnostics, treatment planning, and healthcare operations by evaluating complex datasets including patient records, genetic markers, and predictive models (Hinostroza Fuentes et al. 2025; Freyer et al. 2025; Dietrich 2025). As Truhn et al. (2026, p. 11) puts it: "these systems might independently order diagnostic tests or adjust medication dosages. This represents a fundamental shift from assistive to collaborative intelligence, in which AI agents become active participants in clinical and research workflows rather than merely consultative tools". In simple terms, agentic AI consists in a decentralized computational network of AI agents that together execute operations by themselves. In that sense, the network acts, not merely informs.

The integration of multiple LLMs into an agentic AI enables dealing with vast amounts of unstructured data, accessing databases, navigation of interfaces or the creation of codes. A key function of agentic AI thus lies in the decomposition of tasks into architectures with an orchestration framework, often with a primary agent and several subspecialized support-agents (Moritz et al. 2025; Sapkota et al. 2026). Multi-step reasoning and planning mechanisms facilitate a dynamic sequencing of subtasks, also allowing the system to adapt to environmental changes. Agentic AI is further characterized by the fact that "reflective reasoning and memory systems allow agents to store context across multiple interactions, evaluate past decisions, and iteratively refine their strategies" (Sapkota et al. 2026, p. 7). Natural language interfaces are developed for communication between the AI agents. Using human-like language allows for a flexible and adaptable communication within the system and, at the same time, increases the interpretability and transparency of the decisions as it can be audited and reviewed by humans (Moritz et al. 2025).

Agentic AI is currently being explored across a range of health-related contexts, from basic research in biomedicine (Gao et al. 2024) to envisioned applications in clinical care (Collaco et al. 2025; Gorenshtein et al. 2025). The concrete application of agentic systems has already been researched in exploratory studies for various tasks such as evidence synthesis, clinical documentation, virtual care assistants, predictive and risk analyses or to support diagnostic reasoning and treatment planning (Gorenshtein et al. 2025). These systems can engage with human stakeholders (e.g. healthcare professionals or researchers) or operate autonomously, with inter-agent communication occurring in the background and accessed primarily for oversight or transparency. In medicine, some agentic systems are designed as clinician-facing tools for information retrieval or care coordination; others interact directly with patients (Heydari et al. 2025), for instance to inquire about preferences or obtain consent for diagnostic or therapeutic steps. An additional category comprises systems that simultaneously engage both stakeholders within the same clinical pathway (Campos et al. 2025). These developments could elevate medical AI to the status of "teammates to human clinicians" (Zou and Topol 2025, p. 457) or health companions for patients.

As an example, Moritz et al. present the hypothetical case of a patient complaining of pain in her pelvis who is served by an agentic network of artificial health care providers (Moritz et al. 2025). In that network the main task of providing medical diagnostics and therapy is decomposed and distributed to a number of agents taking care, for example, of symptoms and wearable data ("General Practitioner"), surgery, radiology, pathology, but also of care coordination and scheduling as well as advice from the healthcare insurance. Agentic AI in healthcare therefore might include agents that are directly involved in patient care as well as supportive systems operating at the administrative and management levels.

As with forms of medical or generative AI in general (Wang et al. 2023; Haltaufderheide and Ranisch 2024; de Vere Hunt et al. 2025), the deployment of such systems promises significant benefits, but also raises familiar ethical concerns, ranging from safety and accountability to fairness and privacy. However, rather than focusing on these established issues of bioethical oversight, this article shifts attention to a different set of concerns: the potential *moral consequences* (i.e. effects on the medical morality) that agentic AI might bring about within clinical practice – particularly consequences affecting the patient-physician relationship and the normative structures surrounding it. These consequences will be explored through the lens of *techno-moral change*, a framework that traces how technologies can reshape moral beliefs, values, and social practices over time (Danaher and Sætra 2023; Kudina and Verbeek 2019).

This article takes an *anticipatory approach to ethics* to explore how the moral fabric of healthcare may develop in response to agentic AI. At present, there are primarily programmatic debates and only a few examples of – more or less autonomous – agentic networks tested in healthcare settings (Gorenshtein et al. 2025). Most likely, however, the introduction of such technologies will develop rapidly and will have a profound influence on both patient expectations and the self-perception of healthcare professionals. The article adds in a complementary way to ongoing debates on the morally relevant consequences of medical chatbots (Rudra et al. 2026; J.W. Allen et al. 2024; Zohny et al. 2025; Haltaufderheide and Ranisch 2023), avatars (Sestino and D'Angelo 2023), and other technologies that mimic patient-provider interaction (Meadi et al. 2025). However, it takes seriously the transformative role that (semi-)autonomous AI agents may have when applied in healthcare.

The article proceeds as follows: After a brief introduction on agentic AI in medicine, we continue by clarifying the meaning of *agency* and *autonomy* in AI systems. We then further explain our methodological framework in distinguishing between morally relevant consequences and moral consequences of agentic AI. In the subsequent sections, we spell out possible moral consequences of agentic AI and sketch potential techno-moral changes across three key domains (decisional, relational, and perceptual). We conclude with a discussion of how techno-moral change might be ethically evaluated and addressed through design choices and stakeholder involvement.

## 2. AI agency and autonomy

From a philosophical point of view, the concept of "agency" as inherent in agentic AI deserves further specification. According to a standard definition, agency means an entity's capacity to act intentionally (Schlosser 2019). Intentionality, again, is often linked to causation by the agent's mental states and events. Being an agent would then consist in the ability to perceive and act on an environment in a goal-directed and autonomous way (Russell and Norvig 2016). There are approaches in the philosophy of action, however, holding that intentions cannot be reduced to the classical belief-desire-model of practical reasoning (Hume 1975) that is reliant on the existence of mental states. The reference to mental states – instead of observable agent-like behavior –, in such approaches, is even seen as a shortcoming especially in dealing with non-living entities such as AI systems (Popa 2021).

There are different attempts to dealing with such shortcomings. In synthesizing various aspects of agency into an "agency profile" Leonard Dung, for example, presents a multidimensional approach that aims also to capture the specific nature of *artificial* agency (Dung 2025). Goal-directedness and autonomy, in the sense of functioning without intervention by an outsider, are among the criteria that contribute to the agency profile of an entity. In addition, Dung lists efficacy, planning and intentionality as further dimensions. Intentionality, understood as the ability to act for reasons, is attributed to AI by some authors who eschew the belief-desire-model that refers to mental states, which cannot be easily attributed to machines. Floridi and Sanders (2004), in addition, list interactivity, autonomy and adaptability ("learning") as three essential features for an AI to be considered an agent.

In general, the question whether agency requires phenomenal consciousness is related to fundamental assumptions that are subject to ongoing philosophical debates. The idea of an agency profile or continuum allowing for various degrees of agency is frequently defended together with the use of the word "agenticness" locating agency along a spectrum rather than seeing it as a binary property (Wissuchek and Zschech 2025). Speaking of "agentic" instead of "agent" therefore allows for a more flexible use of the term accounting for distinct properties related to each artificial system.

A second key term is the concept of autonomy as attributed to artificial systems. With respect to AI agents, autonomy refers to "the ability to function with minimal or no human intervention after deployment." (Sapkota et al. 2026, p. 9) This means that autonomous coordinated systems are able to process environmental inputs, reason over contextual data and execute not only predefined, but also adaptive actions in real-time.

Agentic systems that are envisaged for healthcare so far exhibit different degrees to which they are autonomous or goal-directed without human supervision and control. Decisions about human-in-the-loop (participatory) or human-on-the-loop (supervisory) forms of oversight need to be adjusted not only to technological opportunities and reliability but also to risks associated with the employment of agentic AI in healthcare. Workflows operating fully independent of human professionals will not be appropriate in many healthcare settings. In this context, Moritz et al. (2025) refer to complex cases, rare diseases and situations requiring nuanced clinical judgement that might still require human agency. In these situations, agentic AI may still be used in the sense of a supportive tool comparable to conventional AI-driven clinical decision support systems.

## 3. Morally relevant consequences and moral consequences of agentic AI

The integration of agentic AI into healthcare settings entails a range of morally relevant consequences, many of which echo issues already discussed in the context of generative medical AI (Wang et al. 2023; Haltaufderheide and Ranisch 2024; de Vere Hunt et al. 2025), such as hallucinations (Omar et al. 2025; Handler et al. 2025) or biases (Hofmann 2025; Hasanzadeh et al. 2025). Yet the highly automatic (or "autonomous") nature of goal attainment that is inherent to agentic AI leads to a specific risk profile distinguishing it from conventional systems such as LLMs or Large Vision Models (LVMs). For example, cases have been described where AI agents modified their environment to achieve their goal – even engaging in unforeseen behavior, such as rewriting code to bypass task constraints (Gabriel et al. 2025). In a domain as sensitive as healthcare, such goal-driven "shortcuts" pose serious risks to patient safety.

Other specific risks emerge in multi-agent systems operating with limited human oversight. When AI agents interact sequentially, the output of one agent serves as the input of another, which can lead to compounded errors. The performance and reliability of the entire system is then limited by its weakest link ("error propagation") (Moritz et al. 2025; Freyer et al. 2025). Equally pressing are questions of responsibility and liability, as assigning accountability for errors becomes more complex when (semi-)autonomous AI agents contribute to medical decision-making (Gabriel et al. 2025; Lange et al. 2025; Truhn et al. 2026). Moreover, the dynamic behavior and interaction of multiple semi-autonomous agents make their actions difficult to predict and medical AI systems harder to certify (Freyer et al. 2025). Additional threats arise from synthetic requests by adversaries aiming to access sensitive information. For agentic AI to function effectively, it must access diverse data sources and interact across digital ecosystems. This required openness creates vulnerabilities, such as data breaches or unauthorized surveillance, which threaten patient privacy. Especially in healthcare, high degrees of autonomy together with goal-directedness as a functioning principle might lead the agentic AI to execute tasks that are not in line with patient welfare and autonomy as well as with healthcare professionals' ethos.

While the debate on these morally relevant consequences of AI agents is still emerging (Gabriel et al. 2025), many of the issues resemble those known from familiar discussions on medical AI. At the same time, however, agentic AI also triggers *moral consequences*, i.e., it affects moral systems themselves. According to Swierstra et al. (2009), moral consequences describe how technologies reshape moral beliefs, standards, and practices. These changes extend beyond conventional ethical questions. They do not merely arise within an existing normative landscape, rather, they shift its coordinates, altering moral practices, expectations, and values. Historical precedents such as mechanical ventilation, organ and transplantation show how technological innovations can reshape understandings of life, death, and professional responsibility (Nickel 2020). In addition, technologies in health care are known to have unintended consequences, that can also cause changes in social systems (Harrison et al. 2007). Given its active and autonomous role in medical contexts governed by long-standing normative codes, agentic AI may have similar transformative potential.

These moral consequences of evolving AI systems can be examined through the lens of *techno-moral change* (TMC). TMC is a descriptive thesis holding that technology transforms how values and rights are understood while remaining agnostic regarding moral truth (Danaher and Sætra 2023). TMC is not a single theory, but a conceptual framework used to explain and categorize moral changes brought about by technological innovation. Scholars have proposed different lenses for its analysis: In a fundamental way, technology just like other developments can trigger processes of moralization or demoralization (Hopster 2022). With a different perspective, de Boer and Kudina (2021) identify epistemic, existential, and legal dimensions to describe how medical AI may alter normative structures and self-understanding within healthcare. Taking a broader perspective, Danaher and Sætra propose three overarching domains in which technologies can impact moral beliefs and practices: "technology can change morality by changing how we make decisions, how we relate to others and how we understand and perceive the world." (Danaher and Sætra 2023, p. 779)

## 4. Techno-moral change and the patient-physician relationship

Although TMC has become a prominent topic in the philosophy of technology, its potential impact on the moral fabric of medicine remains underexplored, with only a few notable exceptions (de Boer and Kudina 2021). While technological change can affect various moral domains and institutional structures, the patient-physician relationship is a particularly rich site of analysis. As the normative core of clinical practice, it is especially sensitive to being reshaped by the introduction of agentic AI systems.

In general, the patient-physician relationship develops against the background of a complex normative framework that is shaped by ethical and legal provisions as well as historical circumstances. Whereas some aspects of the relationship remain historically rather constant due to the genuine conditions of seeking and providing health care other facets of the relationship have changed significantly. The most impressive example for such a normative shift is probably the rise of patient autonomy and the overcoming of medical paternalism through the second half of the 20th century (Faden and Beauchamp 1986). The paradigmatic situation of an individual in a vulnerable situation encountering a medical expert, by contrast, is still valid in 21st century healthcare. Power dynamics and asymmetries in this encounter necessitate the upholding of key ethical principles such as confidentiality and trust (Pilnick and Dingwall 2011). In addition, patient-centered and empathetic communication serve as prerequisites for providing tailored and effective healthcare to the individual. This also includes healthcare professionals' competency of dealing with medical ignorance and uncertainty and their ability to communicate them to the patient (Villiger 2025).

The introduction of AI technologies has already sparked a debate over the future of the patient-physician relationship (Mittelstadt 2021). Surveys suggest that many clinicians worry AI could undermine their connection with patients (AMA 2023). Scholars have expressed divergent views: some argue AI inherently distorts this relationship (Karches 2018), while others suggest it will not fundamentally alter – or might even replace – traditional forms of medical interaction (Ahuja 2019). However, the dyadic structure of help-seeker and healthcare provider is significantly influenced by the introduction of powerful AI tools as "third partners". Studies indicating the superiority of AI-driven decision-support systems compared to doctors in concrete diagnostic tasks (Liu et al. 2025), for example, unsettle the image of human healthcare providers as authoritative experts. Instead, medical evaluation and advice are now provided by both, human and machine experts with meaningful ethical consequences in terms of authority, responsibility and trust (Funer and Wiesing 2024). Patients increasingly serve as "epistemic partners" in the interpretation of AI outputs that concern their personal health (Salloch and Eriksen 2024). With the introduction of agentic AI in healthcare additional dynamics in the structure of the patient-physician relationship are to be anticipated: Not only human-machine-interaction but also the interaction between a certain number of artificial agents needs to be scrutinized from an ethical perspective. In addition, goal-directedness and high degrees of autonomy as key features of agentic systems bear implications for moral action in healthcare.

Before examining these dynamics more closely, a clarifying methodological remark is necessary: Most existing work on TMC is *retrospective*, drawing on historical cases of moral transformation. One frequently cited example is the transformation of sexual morality following the introduction of pharmacological contraception (Nickel et al. 2022). By examining social and cultural developments *after* the diffusion of such technologies, these analyses show how norms and values were reshaped through their adoption and use.

In contrast, the present article adopts a *prospective* orientation. Rather than diagnosing moral change after the fact, we aim to anticipate plausible trajectories of medical morality associated with the potential deployment of agentic AI. This aligns our approach with various strands of anticipatory ethics (Boenink et al. 2010; Brey 2012; Floridi and Strait 2020; van de Poel 2021; Umbrello et al. 2023), which seek to assess the ethical implications of emerging technologies before their full integration into society.

Of course, such forward-looking assessments are inherently challenging and intensify familiar dilemmas of anticipatory technology assessment. On the one hand, the effects of a new technology often cannot be fully known before widespread implementation; on the other hand, once those effects become visible, the technology may be too entrenched to allow for meaningful intervention (Collingridge 1980). The lens of TMC adds another layer of complexity (Kudina and Verbeek 2019). Not only are technological consequences uncertain, but these very consequences may also shift the moral landscape of future medicine. Thus, introducing uncertainties that can also affect anticipatory ethical assessment itself.

Despite these uncertainties, prospective ethical inquiry need not be arbitrary. Although speculative in nature, it is a well-established in technology assessment. The goal is not prediction but the identification of realistic possibilities (Hopster 2022). Moral change, while unpredictable in detail, follows patterns and operates within logical and conceptual constraints. Our goal is therefore to offer "informed speculation" (Danaher 2021): projections that gain plausibility from various sources such as historical analogies and conceptual analysis.

Comparisons with earlier shifts in medical morality – especially those triggered by existing medical AI – can help illuminate how present trends may continue as AI systems become more autonomous. At the same time, conceptual analysis of core ethical values helps map a "logical space of variation" (Danaher 2021). According to this view, a multiplicity of values necessitates structural trade-offs – for example, between maximizing individual welfare and ensuring justice – which no technology can resolve. What technologies can do, however, is to shift how such trade-offs are framed or which values are prioritized, but not to the resolution of unconcealable options.

In what follows, we draw on a taxonomy of TMC developed by Danaher and Sætra (2023) to analyze how agentic AI may affect and transform the doctor-patient relationship. We structure this discussion along their three domains of TMC: decisional, relational, and perceptual.

## 4.1. Decisional domain: changing how morally charged decisions are made

The introduction of agentic AI into healthcare has the potential to transform the decisional landscape of clinical practice and, with it, the normative structure of the patient-physician relationship. In the framework of TMC presented by Danaher and Sætra (2023), this involves altering option sets, by introducing new possibilities or removing existing ones, and reshaping decision-making costs, e.g. by making certain actions more or less burdensome.

### 4.1.1. New options and obligations

Advances in medical AI already offer clear examples of how technology reshapes available options and associated responsibilities. As powerful diagnostic tools and clinical decision-support systems become more widespread, they do not merely add to the list of possibilities, they can "kiss to life dormant obligations and responsibilities by supplying new 'cans' which result in new 'oughts'" (Swierstra et al. 2009, p. 133). Some have therefore begun to argue that a "professional obligation to use AI systems" may emerge in light of their growing capabilities (Hoffman et al. 2025). At the same time, AI-based practices may render previous standards for diagnosis or treatment inadequate towards the patients and thus morally questionable.

Agentic AI systems are likely to broaden these normative shifts. If such systems demonstrate superior performance in areas like diagnostics at some point in time, failing to use them might come to be seen as professional negligence, a tendency that would reshape what counts as responsible care. Moreover, these changes extend beyond the obligations of professionals. For patients, agentic AI, especially in the form of personal health agents (Heydari et al. 2025), expands their option set by enabling interaction with autonomous systems instead of relying solely on clinicians. This could introduce new forms of access to care, such as 24/7 medical advice or counselling, effectively creating new routes for healthcare provision.

### 4.1.2. New costs

These technological advances, however, often come with decision-making costs for physicians and patients. In the context of "conventional" medical AI, we can already observe ambiguous impacts on medical practice. On the one hand, AI can streamline decisions or automate routine tasks, raising hopes that it might free up physicians' time for interacting with patients (Topol 2019). Similar expectations have been voiced for agentic AI, which is anticipated to "give physicians back valuable time" in the future (Truhn et al. 2026). On the other hand, the digital transformation of healthcare creates new costs for physicians ranging from an increasing need for digital literacy and added "data work" (Morley and Floridi 2025) to new "duties of explainability" (de Boer and Kudina 2021) toward patients, especially when AI outputs are incomprehensible or even opaque. At the same time, possible deskilling effects of physicians are already documented (Budzyń et al. 2025), which are likely to become more pronounced with the increasing autonomy of AI agents (Truhn et al. 2026; Tripathi et al. 2026).

The architectures and complexity of interacting agents could create additional burdens for clinicians. Unlike conventional AI systems in medicine, agentic AI does not only act autonomously, it will likely be broad in application and highly adaptive (Freyer et al. 2025). This would require physicians to pay increased attention to these systems (Sparrow et al. 2025), increasing cognitive labor. Given the goal-oriented functioning of agentic networks, physician oversight may come to resemble the supervision of junior colleagues or entire teams, rather than the simpler oversight required for "conventional" medical AI tools. This also intensifies concerns about physician replacement. Moreover, it could complicate shared decision-making, as both doctors and patients must assess whether complex tasks should be delegated to an agentic AI network. This would likely increase the complexity of shared decision-making, as doctors and patients need to make decisions on whether or not to entrust complex tasks to an agentic network. Moreover, it is likely to increase the complexity and cost of shared decision-making, as doctors and patients must not only decide whether to entrust complex tasks to an agentic network, but also reconcile potentially diverging recommendations from human clinicians and AI agents.

For patients, additional trade-offs may arise from agentic AI systems that shift decision-making costs. The presence of effective agentic AI in clinical practice raises the need to weigh up the protection of their private health information. This includes not only standard medical records, but also health-related data harvested from mobile health tools and personal devices. To allow AI agents to function effectively, sensitive information might need to be disclosed and distributed through networks of AI agents (Liu et al. 2025). Integrating AI into healthcare infrastructures often demands extensive data access, potentially eroding established expectations of confidentiality (Dennstädt et al. 2025). To benefit from agentic AI, patients may find themselves bargaining their privacy for effective healthcare services.

Furthermore, patients may also likely face new costs concerning accountability and transparency. If AI agents operate in ways that are partially opaque, even to the physicians using them, this may require an attitude of "Don't ask me how it works, but it works" (M.R. Allen et al. 2024), which could gradually weaken explainability standards in clinical care. The introduction of agentic AI, with its autonomous and distributed features, may render traditional notions of responsibility increasingly difficult to apply – if not entirely misplaced (Acharya et al. 2025) – since no single human or artificial entity can be clearly held accountable (Truhn et al. 2026). Patients could be forced to trade the possibility of attributing responsibility for the benefits of using highly effective systems – leaving "no one" clearly accountable in the case of failure.

## 4.2. Relational domain: changing how we relate to others

In the framework of TMC, the relational domain concerns how technologies enable new relationships, alter expectations and burdens within existing ones, and shift underlying power dynamics between actors and institutions (Danaher and Sætra 2023). This is particularly relevant for healthcare, where the patient-physician relationship is not only a cornerstone of medical practice and professional ethos but also an important factor influencing clinical outcomes.

### 4.2.1. New relationships

The introduction of digital technologies and AI into healthcare has long been described as a transformative force for the traditional dyadic relationship between patient and physician (Mittelstadt 2021). AI tools can create a *triadic* relationship between physician-patient-AI, or even two distinct dyads: one between physician and AI, the other between patient and AI (Sparrow et al. 2025). Generative AI in medicine, in particular, is increasingly described as a "third actor" in clinical settings (Campos et al. 2025), which may further transform relationships in medicine and potentially "diminish the clinician's role in the decision-making process" (Morley and Floridi 2025).

With the use of agentic AI, new relational constellations are likely to emerge. Agentic AI systems will amplify the mediating role of AI in healthcare by expanding their function beyond decision-support to more autonomous and interactive forms. Their conversational capabilities, especially those leveraging LLMs and multimodal interfaces, make it increasingly plausible that such systems will be perceived not merely as tools, but as interaction partners. A personalized agent might even serve to mirror the human in the system and function as the patient's trusted intermediary over time. Some envision agentic AI as personalized health assistant, i.e. as "lifelong, personalized health agents acting as a co-pilot for users' daily experience" (Heydari et al. 2025), which could foster strong relational bonds between patients and artificial agents.

Unfamiliar relational constellations may arise through multi-agent systems. The triadic model could evolve into a *polyadic* relationship involving the patient, the clinician, and multiple AI agents – even including relationships among AI agents themselves. For example, some proposals suggest that competing AI agents could be used to optimize outcomes (Heydari et al. 2025). More generally, the digital ecosystems from which AI agents originate (e.g. clinics, scientific databases, insurance companies) may exert significant influence on the performance and outputs of the overall agentic network. In such configurations, coordination is typically managed by an AI orchestrator, which may assume a dominant role in mediating conflicts and aligning the activities of the artificial agents. This raises the prospect that relational dynamics in healthcare could increasingly be shaped by background computational infrastructures rather than by direct human interaction.

### 4.2.2. Expectations

Not only can agentic AI introduce new constellations in medicine, it may also reshape expectations and relational norms that govern existing ones. With the rise of agentic AI systems, patients may increasingly expect physicians to act as explainers or guarantors of AI outputs. In clinical settings where agentic AI can produce complex or opaque recommendations, patients might look to their doctors to interpret and justify these outcomes, which effectively turns clinicians into the human face of autonomous systems.

At the same time, concerns are already being raised that AI in medicine may warp patient expectations (M.R. Allen et al. 2024). The overly helpful and often empathic tone of current conversational agents has led some physicians to reflect critically on their own bedside manner (Reisman 2024). As patients increasingly engage with agentic AI systems, they may begin to expect similar levels of communicative polish and responsiveness from their human caregivers – expectations that could put additional pressure on clinicians to match the consistency and availability of these systems (Earp et al. 2025).

An aggravation of such effects can be anticipated with the use of agentic AI. For example, LLMs are often overly confident and optimized for helpfulness. This can lead to the amplification of confirmation bias (Lopez-Lopez et al. 2025) and a sycophantic tendency to prioritize agreement over accuracy (Chen et al. 2025). Although this may depend on future training of AI agents, such tendencies could shift expectations within the patient-physician relationship: eroding the perceived legitimacy of epistemic humility and increasing pressure on physicians to present themselves with equal certainty and optimism.

### 4.2.3. Power dynamics

Relational changes induced by technologies can also entail shifts in power dynamics. Digital health technologies have long been associated with a promise of patient empowerment (Topol 2015), and generative AI expands on this promise. By providing tailored health information, conversational agents can enhance health literacy and help patients enter clinical encounters better informed (Hryciw et al. 2023). Agentic AI could broaden this effect even further. Especially when deployed as personal health assistants, these systems may reduce the traditional epistemic asymmetry between patients and physicians, allowing for more collaborative and potentially more equitable clinical interactions.

However, this redistribution of power is not without complications. While informed patients may engage more confidently in shared decision-making, the increasing authority attributed to autonomous AI systems may also challenge the physician's role, not only in terms of knowledge, but also in terms of moral and professional legitimacy (see below). The consequences of this shift will likely depend on how agentic AI systems are integrated and how their authority is communicated and interpreted within healthcare settings. Furthermore, dealing with power dynamics among artificial agents themselves constitutes a new and potentially highly important technological and ethical challenge to ensure their functioning in critical health care contexts.

## 4.3. Perceptual domain: changing how we perceive and understand moral situations

Technologies do not merely alter the options and relationships through which medical morality unfolds, they also shape how we perceive, interpret, and assign meaning to morally salient aspects and core concepts of medicine. In the framework of TMC, the perceptual domain refers to how technologies introduce new forms of information, metaphors, and conceptual framings that influence how moral agents understand their environment (Danaher and Sætra 2023). In the context of agentic AI in healthcare, such transformations are likely to be subtle yet far-reaching, affecting concepts of health and disease, the clinical realm, or what counts as health-relevant information.

### 4.3.1. Medicalization

Already with the increasing use of data-driven technologies in healthcare, profound changes to our understanding of health, disease, and the nature of the patient have been noted. It has been argued that medical AI reinforces a mechanistic view of medicine. Increased reliance on algorithmic systems can promote a conception of human beings as entities comprehensible through programmable laws and computational models (Karches 2018). This tendency may intensify with the deployment of agentic AI systems, which are likely to operate across multiple contexts and draw on a vast array of patient data. Current visions for such systems imagine them continuously collecting and interpreting data from a wide range of sources, including everyday digital behavior (Capulli et al. 2025). This could blur the boundary between healthcare and daily life and change the perception of what counts as medical data, what is considered clinically relevant, and what defines the medical realm (Haltaufderheide et al. 2025). Such developments may foster a continuous, pervasive form of health monitoring that reconceptualizes health as something to be constantly optimized through technology, rather than episodically managed in clinical encounters.

### 4.3.2. Empathy

Agentic AI may also alter how we perceive key elements of care relationships, such as empathy, as well as how we understand the medical profession and the value of particular tasks. One of the surprising findings from recent research is that users sometimes perceive generative AI models as more empathic than human physicians (Ayers et al. 2023). AI agents, particularly those designed for patient interaction, are likely to deepen this phenomenon (Moritz et al. 2025). Long-term engagement with such agents, especially when reinforced through feedback loops, could foster emotional attachment (Gabriel et al. 2025). These effects may be amplified through anthropomorphic or avatar-based designs.

If interactions with AI systems are perceived as equally or even more emotionally satisfying than those with human caregivers, this could reshape how empathy and care are understood in clinical practice (Karches 2018). Accepting that patients prefer simulated empathy from AI agents could thus transform the privileged moral status of genuine human connection, replacing it with functional relationships to machines (Kreitmair 2025). Empathy risks being reframed as a technical performance rather than a human capacity. Over time, this could recalibrate how both patients and professionals value human presence in healthcare.

### 4.3.3. Medical authority

At the same time, this may also shift perceptions of the importance of other clinical tasks and professional values. While human connection has traditionally been seen as crucial for therapeutic outcomes – especially in psychotherapy – the realization that such relationships can be simulated or substituted by AI agents could lead to a devaluation of the professional labor involved. As De Boer and Kudina (2021) maintain, when medical tasks are delegated to machines, their perceived importance may diminish, because the task itself becomes reframed as technically trivial.

Alongside such shifts, the rise of agentic AI systems that act autonomously also has the potential to transform how physicians are perceived in their professional role. Despite efforts to promote shared decision-making, the clinical encounter has long been structured around the physician's exceptional status as a trained medical expert and gatekeeper of treatment decisions. This perception may erode if agentic AI is increasingly seen as (more) accurate, reliable, or even better at explaining medical content. Over time, this could challenge not only the physician's epistemic authority but – in the case of agentic AI – even the long-standing principle of professional exclusivity in therapeutic decision-making. The introduction of autonomous, decision-capable AI agents therefore may destabilize established ideas about who retains authoritative control over clinical decisions (Morley and Floridi 2025). In this light, the medical profession may no longer appear as the sole or final arbiter of therapeutic pathways, but as one voice among others in a technologically pluralized medical landscape.

# 5. Agentic AI and the ethical transformation of the patient-physician relationship

As indicated, TMC is first and foremost a descriptive thesis, which we have applied in an anticipatory fashion to the prospective introduction of agentic AI in healthcare. Our aim was to identify plausible trajectories of moral transformation in the patient-physician relationship, while acknowledging the uncertainties inherent in evolving social and technological systems that co-shape one another.

Yet this descriptive analysis can feed into a normative perspective. If agentic AI is likely to transform moral attitudes and "lived morality" in clinical practice, such changes cannot only be identified but can also be assessed ethically. At this stage, such normative considerations can only be sketched in outline. Future assessments must be supported by empirical insights into operational practices and stakeholder perspectives as the real-world deployment of agentic AI continues.

Viewing technological innovations in healthcare like agentic AI through the lens of TMC reminds us that even established moral norms, such as those governing medical practice, are historically contingent and susceptible to technological influence. This insight has two important implications: First, it compels us to reflect on which moral norms in medicine might require protection, and which changes may be welcomed. Second, it invites scrutiny how technological trajectories can be steered in ways that support those values we consider essential, or whether certain technologies or technological features should be introduced at all.

Turning to the first dimension, we must ask what we value in medical morality. Even long-standing norms, such as those shaping the patient-physician relationship, are historically contingent and open to transformation. This raises the question of which moral norms ought to be preserved and which changes may be welcomed. Such questions inevitably lead to normatively charged answers and thus go beyond the descriptive scope of TMC. The shift from description to the prescriptive task of deciding which norms can be left to the contingencies of TMC presents its own challenge – most notably, the lack of a stable normative reference point. A helpful point of orientation is to shift attention to the foundational purpose of medicine itself. One prominent account understands medicine as a "healing relationship" (Pellegrino 2006), a practice oriented toward the promotion, restoration, and preservation of health. On this view, distinctive features of medical morality, such as trust, confidentiality, and shared decision-making, derive their moral force not (only) from tradition but gain necessity because they help to achieve these goals.

If the characteristics of the patient-physician relationship as we know it today are not understood as ends in themselves but as means to achieve health-related outcomes, it may be justifiable to reconfigure some of its normative parameters. This goal-oriented view on medical morality creates space for transformation and for welcoming certain TMCs. For instance, if polyadic constellations involving patients, clinicians, and AI systems can yield equally effective or better outcomes, they may be ethically acceptable – even if they depart from familiar modes of clinical interaction. Likewise, some features of existing professional roles or concepts like patient autonomy may be open to reinterpretation, provided that core goals of health and well-being are respected.

Turning to the second dimension, a focus on technology itself, reminds us that not all change is equally controllable, or equally acceptable. Here, it is helpful to distinguish between two types of TMC. Some TMCs are *inherent* to a technology's features and function, while others are *empirically contingent*, depending on context, implementation, and interpretation. For example, the adoption of opaque AI systems in clinical decision-making may inherently be incompatible with established moral practice of obtaining meaningful informed consent. Here the lack of transparency would in itself undermine justified expectations on the comprehensibility. In contrast, a more *contingent* change might involve shifting standards for professional error. If high-performance AI diagnostics lead society to view human error as increasingly unacceptable, this could raise expectations for clinicians. Alternatively, if both human and AI fallibility are acknowledged, tolerance for error might broaden. Such outcomes are not predetermined but rather highlight the probabilistic and path-dependent nature of possible TMC.

These distinctions matter because they inform how much room there is for intervention through design choices of technology. For contingent effects, careful design and implementation can help align technological development with established medical morality or to deliberately steer morality into desirable directions. For inherent effects, stronger scrutiny may be needed. In some cases, this may even lead to questioning whether certain technologies or features should be introduced at all, particularly in sensitive domains of healthcare. For example, the possibility that agentic AI might erode accountability through distributed autonomy has already been cited as an argument against pursuing such systems (Mitchell et al. 2025).

The longstanding role of moral norms in medicine further supports such a cautious approach. When core values of current patient-physician relationship such as confidentiality, trust, and responsibility are at stake, it is prudent to prioritize design choices that allow for reversibility or adaptation. This is particularly important given the experimental character of innovative technologies such as agentic AI (Ranisch and Haltaufderheide 2025). Their full effects on complex social practices such as medical morality cannot be known in advance. Hence an incremental approach seems warranted, which allows for learning mechanisms on technological effects, including on, how they can shift moral frameworks in medicine.

From a more practice-oriented perspective, meaningful involvement of patients and other stakeholders is crucial for the ethical development of experimental technologies like agentic AI that could have unforeseen consequences on the moral fabric of medicine. Participatory processes can help ensure that agentic systems are designed with sensitivity to moral concerns and adapted to diverse healthcare settings. Take, for example, informed consent, a practice likely to remain central for future medical morality. As agentic systems increasingly operate through complex multi-agent interactions, developers must identify appropriate points at which patients are actively engaged and empowered to make meaningful choices. Otherwise, agentic AI may erode standards of patient autonomy, and possibly lead to some form of "computer paternalism" (McDougall 2019) which occurs in new guise when implementing agentic AI. Proper forms of involvement (e.g. through digital interfaces) for various health care-related tasks and groups of patients must be developed considering the needs of, e.g., minors, incompetent patients or individuals with a low health literacy. Such appropriate forms of involvement will help to secure that moral changes in inherent normativity of medicine do not occur without the control and steering of those who are primarily affected.

## 6. Conclusions

The advancement of generative AI has opened new possibilities for medicine and healthcare, ranging from administrative tasks to diagnosis or clinical decision-making. Among the most recent developments is the rise of agentic AI: a class of LLM-based systems that not only provide information but are capable of executing goal-directed tasks with a high degree of autonomy. These technologies are envisioned as teammates to clinicians or as personal companions to patients. While the promises of agentic AI are substantial, so too are the risks.

Much of the current discourse on risks focuses on the "hard" impacts of these technologies: risks related to safety or patient health. However, often overlooked, there are "soft concerns" that may be equally far-reaching (Swierstra and te Molder 2012): subtle but profound effects on the normative foundation of health care. After all, medicine is not merely a technical or operational field, but a social practice, governed by distinguished moral standards, professional codes and long-established norms, especially in the patient-physician relationship.

Given the features of agentic AI – especially its increased autonomy and its capacity to introduce new forms of interaction between patients, clinicians, and artificial agents – it seems likely that these technologies will have an impact on the moral fabric of medicine. Then, agentic AI will not only lead to morally relevant consequences (such as biases or fairness issues), but may also affect medical morality itself by influencing how patients and physicians relate to each other, how their decisions are being made, and how they perceive the medical realm.

Recognizing that agentic AI may shape morality also entails accepting moral responsibility for their design and deployment. AI technologies should therefore be designed and implemented in ways that reinforce the values we have reasons to preserve, while enabling desirable forms of moral change where these are seen as justified. This insight becomes a cornerstone of ethical foresight and design ethics in the context of medical AI, highlighting that design choices are, inherently, normative choices.

This last point, however, leads to a deeper philosophical question: From which perspective should we evaluate changes in medical morality? Eventually, "soft" concerns about value shifts point us back to hard questions of medicine itself. What is the proper aim of healthcare? What makes a clinical relationship good? Addressing these questions remains essential if we are to steer the moral transformations that may accompany agentic AI, without losing sight of the deeper goals defining medicine as a moral practice.

# Disclosure Statement

The authors have no conflict of interest to report.

# Additional information

## Abbreviations

- AI: Artificial Intelligence
- LIM: Large Image Model
- LLM: Large Language Model
- LVM: Large Vision Model
- TMC: Techno-moral change

## Funding

This work was supported in part by the Volkswagen Foundation ("Digital Medical Ethics Network", grant number: 9B233) for R.R. and by the Deutsche Forschungsgemeinschaft (Scientific Network "Digital Bioethics", project number 525059925) for S.S.

## Acknowledgement

ChatGPT 5.2 (OpenAI) and Claude Opus 4.6 (Anthropic) were used to support language editing of parts of the manuscript. All substantive arguments, interpretations, and conclusions were developed by the authors, who take full responsibility for the content.